\documentclass[fleqn,twoside]{article}
\usepackage{espcrc2}
\usepackage{amsmath}
\usepackage{graphicx}

\title{Yukawa Matrix for the Neutrino and Lepton Flavour Violation
\thanks{Presentation given by K.Tsumura}}

\author{
S. Kanemura\address[osaka]{Department of Physics, Osaka University Toyonaka, Osaka 560-0043, Japan},
 K. Matsuda\addressmark[osaka],
 T. Ota\addressmark[osaka],
 T. Shindou\address{Theory group, KEK, Tsukuba 305-0801, Japan},
 E. Takasugi\addressmark[osaka]
 and K. Tsumura\addressmark[osaka]\thanks{ko2@het.phys.sci.osaka-u.ac.jp}
}

\begin{document}

%%%%%%%%%%%%%%%%%%%%%%%%%%%%%%%%%%%%%%%%%%%%%%%%%%%%%%%%%%%%%%%%%%%%%%%%%
\begin{abstract}
We estimate the magnitude of Lepton Flavour Violation (LFV) from the phase 
of the neutrino Yukawa matrix. 
In the minimal supersymmetric standard model with right-handed neutrinos, 
the LFV processes $l_i \to l_j \gamma$ can appear through 
the slepton mixing, which comes from the renormalization group effect 
on the right-handed neutrino Yukawa interaction between 
the Grand Unified Theory scale and the heavy right-handed neutrino mass scale.
Two types of phases exist in the neutrino Yukawa matrix.
One is the Majorana phase, which can change the magnitude of the LFV 
branching ratios by a few factor. 
The other phases relate for the size of the Yukawa hierarchy and its phase 
effect can change the LFV branching ratios by several orders of magnitude.
\end{abstract}
%%%%%%%%%%%%%%%%%%%%%%%%%%%%%%%%%%%%%%%%%%%%%%%%%%%%%%%%%%%%%%%%%%%%%%

\maketitle

%%%%%%%%%%%%%%%%%%%%%%%%%%%%%%%%%%%%%%%%%%%%%%%%%%%%%%%%%%%%%%%%%%%%%%
\section{Introduction}  
Lepton flavour violation (LFV)is an important signature 
into physics beyond the standard model (SM). In the SM with massive 
neutrinos, neutrinos have the Yukawa interaction.
Lepton flavour is violated like quark flavour.
However, LFV is strongly suppressed by 
the neutrino masses. Supersymmetry (SUSY) with massive neutrinos 
makes the situation drastically changed.
It can predict sizable LFV effects, because the alternative source of 
LFV is generated from the slepton mixing through the renormalization group 
effect on the right-handed neutrino Yukawa interaction. 

The LFV processes $l_i \to l_j \gamma (i \ne j)$ of charged lepton are 
being measured. The MEG experiment \cite{MEG} gives 
the upper bound on the $\mu \to e \gamma$ process as
\begin{align}
{\rm Br}(\mu \to e \gamma) \le 1.2 \times 10^{-11}.
\end{align}
The forthcoming experiment reaches to ${\cal O} (10^{-14})$.
The $\tau$ decay processes are also measured at B-factories \cite{Abe:2003sx}. 
On the other hand, it has been already discovered that lepton flavour is 
violated in the neutrino sector. The oscillation parameters will be 
measured in detail.

In this letter, we study the phase of the neutrino Yukawa matrix and 
investigate its effect on the magnitude of the LFV processes.
%%%%%%%%%%%%%%%%%%%%%%%%%%%%%%%%%%%%%%%%%%%%%%%%%%%%%%%%%%%%%%%%%%%%%%

%%%%%%%%%%%%%%%%%%%%%%%%%%%%%%%%%%%%%%%%%%%%%%%%%%%%%%%%%%%%%%%%%%%%%%
\section{The neutrino Yukawa matrix}
In the framework based on the see-saw mechanism, we can parameterize 
the neutrino Yukawa matrix $Y_\nu$ in terms of physical quantities 
as \cite{Pascoli:2003rq}
\begin{align}
\frac{v_u}{\sqrt{2}} Y_\nu = \sqrt{M_R^{diag}} {\cal R} \sqrt{m_\nu^{diag}} U_{MNS}^\dag ,
\end{align}
 where $v_u$ is the vacuum expectation value of the Higgs boson, 
$U_{MNS}$ is the observed Maki-Nakagawa-Sakata (MNS) matrix including two
 Majorana phases $\xi_{1,2}$; i.e.,
\begin{align}
\hspace{-9mm}
U_{MNS} \sim
\begin{pmatrix}
0.85 & -0.53 & 0 \\
0.37 & 0.60 & -0.71 \\
0.37 & 0.60 & 0.71
\end{pmatrix} 
\begin{pmatrix}
1 && \\ &e^{i \xi_1}& \\ &&e^{i \xi_2}
\end{pmatrix}.
\end{align}
Here we neglect the $(1-3)$ element of $U_{MNS}$.
$m_\nu^{diag}$ is the neutrino mass matrix and $M_R^{diag}$ is 
the right-handed Majorana neutrino mass matrix in each diagonal base.
An arbitrary complex orthogonal matrix ${\cal R}$ is expressed as
\begin{align}
{\cal R} \equiv O_{12} O_{23} O_{31} Q_{12} Q_{23} Q_{31},
\end{align}
where
\begin{align} 
\hspace{-8mm}
O_{12}(\theta_{12})\equiv \left( \begin{array}{ccc} \cos \theta_{12} & - \sin \theta_{12} & \\
                                  \sin \theta_{12} & \cos \theta_{12} & \\
                                    & & 1 \end{array} \right), \rm{etc.},
\end{align}
and
\begin{align}
Q_{ij} \equiv O_{ij}(i \eta_{ij}).                                  
\end{align}
The elements $Q_{ij}$ give a drastic effect on structure of the Yukawa 
matrix as hyperbolic functions. We call $i \eta_{ij}$ as {\it R-phase} 
to distinguish it from the Majorana phase. 

In the following, we assume that the neutrino masses are {\it approximately} 
degenerate and the right-handed neutrino masses are degenerate
\begin{align}
&\hspace{-5mm}
m_\nu^{diag} \sim m \left( \begin{array}{ccc}
                           1 & & \\
                           & 1 + \frac{\Delta m^2_\odot}{2 m^2} &\\
                           & & 1 + \frac{\Delta m^2_@}{2 m^2}
\end{array} \right) ,\\
&\hspace{0mm}
M_R^{diag} \sim M_R \left( \begin{array}{ccc}
                           1 & & \\
                           & 1 &\\
                           & & 1
\end{array} \right),
\end{align}
where, $\Delta m^2_{\odot}$ and $\Delta m^2_@$ are the solar and atmospheric 
neutrino mass squared differences, $M_R$ is the size of right-handed neutrinos,
and $m$ is undetermined light neutrino mass parameter.
We then obtain the neutrino Yukawa matrix $Y_\nu$ in its diagonal base as
\begin{align}
Y_\nu^{diag} \sim \frac{\sqrt{2}}{v_u} \sqrt{M_R m}
\left(
\begin{array}{ccc}
1/r & & \\ & 1 & \\ & & r
\end{array}
\right),
\end{align}
where
\begin{align}
&r^2 \equiv 2x^2-1+2x \sqrt{x^2-1}, \label{R1} \\ 
&x \equiv \cosh \eta_{12} \cosh \eta_{23} \cosh \eta_{31}. \label{R2}
\end{align}
This expression indicates the characteristic relation $(y_1/y_2 = y_2/y_3)$ 
among the neutrino Yukawa couplings, which is similar to those of 
the quarks and the charged leptons $(m_u/m_c \sim m_c/m_t, {\rm etc.})$.
For $m \sim 0.1 \rm{eV}$ and $M_R \sim 10^9 \rm{GeV}$, $r$ is close to 
${\cal O}(10^2)$, so that the neutrino Yukawa couplings become hierarchical.
The parameter $r$ is written in terms of the combination of {\it R-phase} 
(see Eqs.(\ref{R1}) and (\ref{R2}) ), 
and it determines the size of the neutrino Yukawa hierarchy.
%%%%%%%%%%%%%%%%%%%%%%%%%%%%%%%%%%%%%%%%%%%%%%%%%%%%%%%%%%%%%%%%%%%%%%

%%%%%%%%%%%%%%%%%%%%%%%%%%%%%%%%%%%%%%%%%%%%%%%%%%%%%%%%%%%%%%%%%%%%%
\section{Lepton Flavour Violation}
We consider the SUSY models, especially the minimal supersymmetric 
standard model (MSSM) with right-handed neutrinos. 
In the slepton sector, the MSSM Lagrangian has an alternative source of 
LFV through the following soft SUSY breaking terms,
\begin{align}
&\hspace{-8mm}- {\cal L}_{soft} 
= \left( A^e_{ij} H_d \tilde{e}_{Ri}^* \tilde{L}_j+ A^\nu_{ij} H_u \tilde{\nu}_{Ri}^* \tilde{L}_j + {\rm h.c.} \right) \nonumber \\
&\hspace{-4mm} + (m_{\tilde{L}}^2)_{ij} \tilde{L}^\dag_i \tilde{L}_j
+ (m_{\tilde{e}}^2)_{ij} \tilde{e}_i^* \tilde{e}_j + (m_{\tilde{\nu}}^2)_{ij} \tilde{\nu_R}_i^* \tilde{\nu_R}_j,
\end{align}
where $A^{e,\nu}$ are the slepton tri-linear couplings and $m_{\tilde{L},\tilde{e},\tilde{\nu}}$ are the soft mass parameters for the sleptons. We assume 
that these couplings are universal at the Grand Unified Theory (GUT) scale 
($M_{GUT}$), i.e.,
\begin{align}
(m_{\tilde{L}}^2)_{ij} &= (m_{\tilde{e}}^2)_{ij} = (m_{\tilde{\nu}}^2)_{ij} = \delta_{ij} m_0^2 \nonumber \\
A^\nu &= Y_\nu a_0 m_0, \, A^e = Y_e a_0 m_0.
\end{align}
In this framework, the LFV processes $l_i \to l_j \gamma (i \ne j)$ can appear 
due to the slepton mixing \cite{Hisano:1995cp}, which comes from the renormalization group effect 
on $Y_\nu$ between the scales of $M_{GUT}$ and $M_R$
\begin{align}
\hspace{-8mm} (\Delta m^2_{\tilde{L}})_{ij} \sim - \frac{1}{16 \pi}(6+2a_0)m_0^2 (Y_\nu^\dag Y_\nu)_{ij} \log \frac{M_{GUT}}{M_R}.
\end{align}
The branching ratios for these processes are expressed by
\begin{align}
{\rm Br}(l_i \to l_j \gamma) \sim \frac{\alpha^3}{G_F^2}
\frac{|(\Delta m^2_{\tilde{L}})_{ij}|}{m_S^8} \tan^2 \beta,
\end{align}
where $m_S$ is the typical SUSY scale, $\tan \beta$ is the ratio of the vacuum 
expectation values of Higgs bosons, $\alpha$ is the fine structure constant, 
and $G_F$ is Fermi constant.

\begin{figure}
\includegraphics[width=7.5cm]{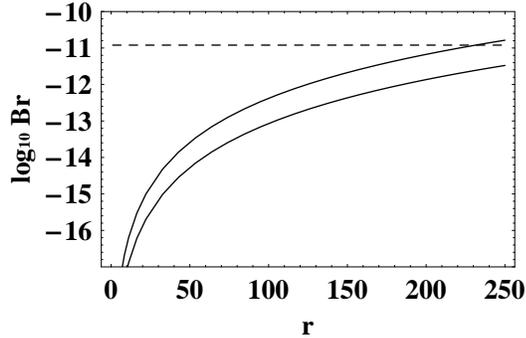}
\vspace{-1cm} 
\caption{The branching ratio of the $\mu \to e \gamma$ 
event for $\alpha=\pi$ (upper solid line) and 
$\alpha=\pi/2$ (lower solid line).
The experimental upper limit from MEG is also shown (dashed line).}
\label{BR-r}  
\end{figure}

Let us consider the $r$ dependence of LFV with $\eta_{23,31}=0$, 
the effect of {\it R-phase} contributes to only $\mu \to e \gamma$ process.
A comparison of each magnitude of LFV processes is already 
meaningless, and we plotted the branching ratio for 
$\mu \to e \gamma $ process (see Figure \ref{BR-r}.),
where we take $m_S = m_0 = 1 {\rm TeV}, a_0=1 , \tan \beta =10 $, 
and $M_{GUT} =  2 \times 10^{16} {\rm GeV}$.
The LFV branching ratio is approximately proportional to $r^4$, 
so that a large value of $r$ can change the branching ratio 
by several orders of magnitude.

\begin{figure}
\includegraphics[width=7.5cm]{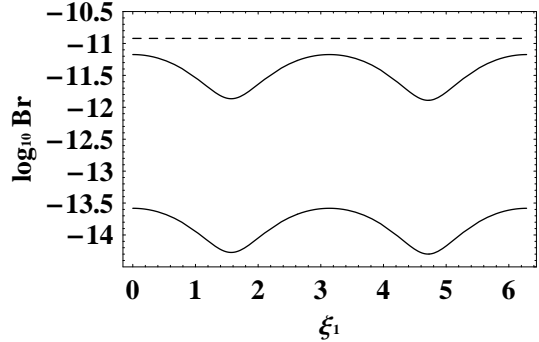}
\vspace{-1cm}
\caption{The branching ratio of the $\mu \to e \gamma$ 
event for $r=200$ (upper solid line) and 
$r=50$ (lower solid line).
The upper limit is also shown (dashed line).}
\label{BR-a}
\end{figure}

On the other hand, we show the dependence on the Majorana phase $\xi_1$
(see Figure \ref{BR-a}.). When $\eta_{23,31}=0$ , the branching ratio 
does not depend on $\xi_2$.
Figure \ref{BR-a} shows that the branching ratio is a periodic function on 
$\xi_1$, and that the effect of Majorana phase is smaller than that of $r$.
%%%%%%%%%%%%%%%%%%%%%%%%%%%%%%%%%%%%%%%%%%%%%%%%%%%%%%%%%%%%%%%%%%%%%%%%%%%%%%%

%%%%%%%%%%%%%%%%%%%%%%%%%%%%%%%%%%%%%%%%%%%%%%%%%%%%%%%%%%%%%%%%%%%%%%%%%%%%%%%
\section{Summary}
We have analyzed the structure of the neutrino Yukawa matrix and have 
discussed the magnitude of LFV processes from the phase effect on 
the neutrino Yukawa matrix in the MSSM with the right-handed neutrinos. 

The neutrino Yukawa matrix has two types of phases, Majorana phases 
and {\it R-phases}. In the case that neutrino masses are degenerate and 
the right-handed neutrino masses are degenerate, the eigenvalues 
of the neutrino Yukawa matrix become hierarchical spectrum and the 
{\it R-phases} determine the size of Yukawa hierarchy.

In the SUSY models, sizable LFV can arise due to the slepton mixing from 
the renormalization group effect on the neutrino Yukawa matrix between 
$M_{GUT}$ and $M_R$.
The Majorana phases can change the LFV branching ratios by a factor, 
and these magnitudes become periodic as the function of Majorana phases. 
On the other hand, {\it R-phases} enhance the magnitude of the 
LFV branching ratios by several orders.
%%%%%%%%%%%%%%%%%%%%%%%%%%%%%%%%%%%%%%%%%%%%%%%%%%%%%%%%%%%%%%%%%%%%%%%%%%%%%%%

%%%%%%%%%%%%%%%%%%%%%%%%%%%%%%%%%%%%%%%%%%%%%%%%%%%%%%%%%%%%%%%%%%%%%%%%%%%%%%%

%%%%%%%%%%%%%%%%%%%%%%%%%%%%%%%%%%%%%%%%%%%%%%%%%%%%%%%%%%%%%%%%%%%%%%%%%%%%%%%

\end{document}